\documentclass[11pt]{article}  
\usepackage{amsmath}  
\usepackage{pstricks}
\usepackage{hyperref} 
 \hypersetup{
     colorlinks=true,
     linkcolor=black,
     filecolor=back,
     citecolor=blue,      
     urlcolor=cyan,
     }
\usepackage[most]{tcolorbox}

\usepackage{tikz} \usetikzlibrary{arrows.meta}
\usepackage[vcentermath]{youngtab}
\usepackage{bbm}
\usepackage{amssymb,amsfonts}

\newcommand{\re}[1] {(\ref{#1})}

\newcommand{\beq}{\begin{equation}}
\newcommand{\eeq}[1]{\label{#1}\end{equation}}

\newfont{\bbbold}{msbm10 scaled \magstep1}

\def\cL{{\cal L}}

\def\cN{{\cal N}}
\def\cO{{\cal O}}

\newfont{\goth}{eufm10 scaled \magstep1}

\def\gl{\mbox{\goth l}}

\def\gs{\mbox{\goth s}}

\def\a{\alpha}
\def\b{\beta}
\def\c{\gamma}
\def\d{\delta}\def\D{\Delta}
\def\e{\epsilon}
\def\f{\phi}

\def\be{\begin{equation}}\def\ee{\end{equation}}
\def\bea{\begin{eqnarray}}\def\eea{\end{eqnarray}}
\def\barr{\begin{array}}\def\earr{\end{array}}

\def\o{\omega}

\let\la=\label

\def\nn{\nonumber}
\def\bd{\begin{document}}
\def\ed{\end{document}}
\def\ba{\begin{array}}
\def\ea{\end{array}}
\def\bea{\begin{eqnarray}}
\def\eea{\end{eqnarray}}
\def\ft#1#2{\tfrac{#1}{#2}}
\def\fft#1#2{\frac{#1}{#2}}
\def\sst#1{{\scriptscriptstyle #1}}
\def\oneone{\rlap 1\mkern4mu{\rm l}}

\newcommand{\eq}[1]{(\ref{#1})}
\newcommand{\w}[1]{\\[0.#1cm]}
\def\eqs#1#2{(\ref{#1}-\ref{#2})}
\def\det{{\rm det\,}}
\def\tr{{\rm tr}}
\def\ad{{\rm ad}}

\newcommand{\hoch}[1]{$\, ^{#1}$}
\newcommand{\imperial}{\it\small Theoretical Physics Group, Imperial College London\\ Prince Consort Road, London SW7 2AZ, UK}
\newcommand{\kings}
{\it\small Department of Mathematics, King's College, University of London\\ Strand, London WC2R 2LS, UK}
\newcommand{\uu}
{\it\small Department of Theoretical Physics, Uppsala, Sweden}
\newcommand{\hip}
{\it\small HIP-Helsinki Institute of Physics, P.O. Box 64 FIN-00014
University of Helsinki, Suomi-Finland}
\newcommand{\stock}
{\it\small Department of Theoretical Physics, Stockholm, Sweden}
\newcommand{\golm}
{\it\small AEI, Max Planck Institut f\"ur Gravitationsphysik\\ Am M\"{u}hlenberg 1, D-14476 Potsdam, Germany}
\makeatletter
\renewcommand\theequation{\thesection.\arabic{equation}}
\@addtoreset{equation}{section} \makeatother

\newcommand{\sa}{/ \hspace{-1.2ex}}
\newcommand{\saa}{/ \hspace{-1.4ex}}
\newcommand{\saaa}{\, / \hspace{-1.6ex}}
\newcommand{\Scal}[1]{\Bigl ({#1} \Bigr )}
\newcommand{\scal}[1]{\bigl ({#1} \bigr )}

\newcommand{\CR}{\nonumber \\*}

\newcommand{\trace}{\hbox {tr}~}
\newcommand{\traceS}{\hbox {tr}_{\scriptscriptstyle \mathfrak{S}}~}

\DeclareMathAlphabet{\mathpzc}{OT1}{pzc}{m}{it}
\def\BRST{\,\mathpzc{s}\,}
\def\aBRST{{\scriptstyle (\mathpzc{s})}}
\def\q{{{\scriptscriptstyle (Q)}}}
\def\qs{{\scriptscriptstyle (Q\mathpzc{s})}}
\def\Qsla{{\mathcal{S}_{\q}}}
\def\Slav{{\mathcal{S}_\aBRST}}
\def\epsilonb{{\overline{\epsilon}}}
\def\bulletup{{\scriptstyle \bullet}}

\newcommand{\gra}[2]{{\scriptscriptstyle (#1 , #2 )}}
\newcommand{\ord}[1]{{\scriptscriptstyle (#1)}}

\newcommand{\na}{\nabla}

\newcommand{\ber}{\begin{eqnarray}}
\newcommand{\eer}[1]{\label{#1}\end{eqnarray}}
\newcommand{\eero}{\end{eqnarray}}

\def\cL{{\cal L}}
\def\cN{\mathcal{N}}
\def\cO{\mathcal{O}}

\def\ie{{\it i.e.}\ }
\def\eg{{\it e.g.}\ }

\newcommand{\sfrac}[2]{{\scriptstyle \frac{#1}{#2}}}
\newcommand{\stfrac}[2]{{\scriptscriptstyle \frac{#1}{#2}}}

 \def\balpha{{\overline{\alpha}}}
 \def\bbeta{{\overline{\beta}}}
 \def\bgamma{{\overline{\gamma}}}
 \def\bdelta{{\overline{\delta}}}
 \def\bepsilon{{\overline{\epsilon}}}
 \def\bvarepsilon{{\overline{\varepsilon}}}
 \def\bzeta{{\overline{\zeta}}}
 \def\bareta{{\overline{\eta}}}
 \def\btheta{{\overline{\theta}}}
 \def\bvartheta{{\overline{\vartheta}}}
 \def\biota{{\overline{\iota}}}
 \def\bkappa{{\overline{\kappa}}}
 \def\blambda{{\overline{\lambda}}}
 \def\bmu{{\overline{\mu}}}
 \def\bnu{{\overline{\nu}}}
 \def\bxi{{\overline{\xi}}}
 \def\bpi{{\overline{\pi}}}
 \def\brho{{\overline{\rho}}}
 \def\bvarrho{{\overline{\varrho}}}
 \def\bsigma{{\overline{\sigma}}}
 \def\bvarsigma{{\overline{\varsigma}}}
 \def\btau{{\overline{\tau}}}
 \def\bphi{{\overline{\phi}}}
 \def\bvarphi{{\overline{\varphi}}}
 \def\bchi{{\overline{\chi}}}
 \def\bpsi{{\overline{\psi}}}
 \def\bomega{{\overline{\omega}}}

\def\thalf{{\textrm{\tiny\textonehalf}}}
\def\tquarter{{\textrm{\tiny\textonequarter}}}
\def\Ko{{\scriptscriptstyle K}}
\def\tKo{\scriptscriptstyle k }
\def\corr{$\clubsuit$} 

\newcommand{\auth}{\large P.S.\ Howe${}^{a,}$\footnote{email: paul.howe@kcl.ac.uk} and U. Lindstr\"om${}^{b,}$\footnote{email: ulf.lindstrom@physics.uu.se}}

\begin{document}

\renewcommand{\thefootnote}{\fnsymbol{footnote}}

\null
\begin{flushright}
{\small UUITP-23/26}
\vskip 1.5 cm
\end{flushright}

\begin{center}
{\Large{\bf Exotic gravity theory in topologically non-trivial spacetimes}}
\vspace{.75cm}

\auth
\end{center}
\vspace{.5cm}

\centerline{${}^a${\it \small Department of Mathematics, King's College London}}
\centerline{{\it \small The Strand, London WC2R 2LS, UK}}
\vspace{.5cm}

\centerline{${}^b${\it \small Department of Physics and Astronomy, Theoretical Physics, Uppsala University}}
\centerline{{\it \small SE-751 20 Uppsala, Sweden }}
\centerline{\it \small and}
\centerline{\it \small Centre for Geometry and Physics, Uppsala University,
SE-75106 Uppsala, Sweden
}

\vspace{1cm}


\centerline{{\bf Abstract}}
\vskip .5cm
An exotic linearised theory of superconformal gravity in $D=6, (4,0)$ superspace has been  proposed by C. Hull. In this note we discuss the bosonic sector of this model in topologically non-trivial spaces. Roughly speaking, Hull's model can be thought of as a square of the $H=dB$ theory, although not all Hull fields can be so presented. We also briefly discuss the form sector of $D=11$  supergravity.
\vspace{1cm}

\renewcommand{\thefootnote}{\arabic{footnote}}
\setcounter{footnote}{0}

\pagebreak
\tableofcontents
\setcounter{page}{1}    


\section{Introduction}  

An exotic theory of linearised superconformal gravity has been proposed by C. Hull \cite{Hull:2000rr, Hull:2000zn}. He discusses it for the maximal case of $D=6, (4,0)$ supersymmetry, although the model exists for any $(p,0)$ with $p\leq 4$ in $D=6$.  The leading bosonic component of the potential multiplet is a field $C_{mn,pq}$ in the $(2, 2)$ Young tableau representation of $\gs\gl(6)$ (i.e. 2 boxes in the first row and 2 in the second), where each pair of indices is taken  to be anti-symmetrised.  
In general we shall refer to fields represented by 2-column Young tableaux as $(p,q)$-forms, generically denoted as $\o_{p,q}$. There are then two exterior derivatives, denoted by $d_1$ and $d_2$ which act on the first and second columns of such forms.
The corresponding multiplet reduces to a standard supergravity multiplet in lower dimensions after dualisation. The theory can be thought of as a square of the $(H,B)$-model, where $H=dB$ is a three-form, which exists for $(p,0)$, $p=1,2$, and which was introduced in \cite{Howe:1983fr}  and \cite{Koller:1982cs}. However, not all Hull potentials will be expressible as products of two $B$ fields in the $(2,2)$ Young tableau representation.  A superspace description of Hull's model has been given by Cederwall \cite{Cederwall:2020dui}, while an action has been given for the theory in a $5+1$ split formulation \cite{Bertrand:2020nob, Bertrand:2022pyi}. The field strength $G_{mnp,qrs}$ is obtained from $C$ by acting with 2 derivatives and is in the $(3,3)$ Young tableau representation. This field is self-dual on both sets of indices, a fact which follows from its presentation as the square of $H$. In a recent paper \cite{Howe:2024ojq} the present authors reinterpreted the bosonic sector of this model in loop space where the double contraction of the loop space version of $C_{2,2}$ with the vector fields generating circle transformations gives rise to a symmetric two-index tensor which can be interpreted as a metric.

In this note we shall be concerned with the bosonic sector of Hull's theory on topologically non-trivial spaces. In the case of the $(H,B)$ system, this has been interpreted in terms of gerbes in the mathematical literature. This shall not be our focus here, since we are interested in bi-forms and their gerbe interpretation seems problematic.

\section{$(B,H)$ model}

In this model, $H=d B$ where $B$ and $H$ are respectively 2 and 3 forms, and we have the descent sequence:

\begin{align}
\label{chainofrelations1} H&= dB_\alpha~,~~~~~&B_\alpha \in \Omega^2(U_\alpha)~,\\
\label{chainofrelations2}  (\d B)_{\a\b}:= B_\alpha - B_\beta &= 
d C_{\alpha \beta}~,~~~~~&C_{\alpha\beta} \in \Omega^1(U_{\alpha\beta})~,\\
\label{chainofrelations3} (\d C)_{\a\b\c}:= C_{\alpha\beta} + C_{\beta\gamma} 
+ C_{\gamma\alpha} &= d D_{\alpha\beta\gamma}~,~~~~~&
D_{\alpha\beta\gamma} \in C^\infty (U_{\alpha\beta \gamma})~,\\
\label{chainofrelations4} (\d D)_{\a\b\c\d}:=D_{\alpha\beta\gamma} 
-D_{\d\a\b} +D_{\c\d\a}-D_{\b\c\d}&= E_{\alpha\beta\gamma\delta}~,~~~~~\\
\end{align}
where the $E_{\a\b\c\d}$ are constants defined on $U_{\a\b\c\d}$ and
where we denote the overlap of $n$ open sets by $U_{\a_1 \a_2...\a_n}$. Here the subscripts refer to the open sets that the fields are defined on with respect to a good open cover $\cal{U}$. 
Thus $B_{\a}$ is defined on each open set $U_{\a}$, $C_{\a\b}$ is defined on each double overlap 
$U_{\a\b}:=U_{\a}\cap U_{\b}$, and so on.

We can be more general and consider a closed $(p+2)$-form $H_{p+2}$ with descendants $B_{(p+1)\a}$, 
$C_{p\a\b}$, etc on each successive overlap. We can also include local gauge transformations $\Delta$
that act on each of the overlaps, $\a, \a\b, \a\b\c$ etc.. For example, $\D B_{p\a}=d b_{(p-1)\a}; \D b_{(p-1)\a}=d b_{(p-2)\a}$ on $U_{\a}$ and so on. In this way we can build up the \v Cech de Rham double complex with overlaps on the horizontal axis and form rank on the vertical.\\

\begin{equation} \label{Hdiagram}
\hspace{-5cm}
\begin{picture}(400,200)
\put(185,180){$B_{p+1}$}
\put(210,182){\vector(1,0){20}}
\put(190,195){\vector(0,1){20}}
\put(130,215){$H_{p+2}$}

\put(160,220){\vector(1,0){20}}

\put(190,155){\vector(0,1){20}}
\put(210,148){\vector(1,0){20}}
\put(235,145){$C_{p}$}
\put(256,148){\vector(1,0){20}}
\put(185,145){$b_{p}$}
\put(190,115){\vector(0,1){20}}
\put(185,100){$b_{p-1}$}
\put(190,73){\vector(0,1){20}}
\put(240,73){\vector(0,1){20}}
\put(280,100){$D_{p-1}$}
\put(185,60){$b_{p-2}$}
\put(240,115){\vector(0,1){20}}
\put(240,155){\vector(0,1){20}}
\put(235,100){$c_{p-1}$}
\put(256,102){\vector(1,0){20}}
\put(240,155){\vector(0,1){20}}
\put(256,62){\vector(1,0){20}}
\put(210,62){\vector(1,0){20}}
\put(256,62){\vector(1,0){20}}

\put(285,73){\vector(0,1){20}}

\put(210,102){\vector(1,0){20}}

\put(285,115){\vector(0,1){20}}
\put(285,35){\vector(0,1){20}}
\put(285,-1){\vector(0,1){20}}
\put(190,35){\vector(0,1){20}}
\put(190,-1){\vector(0,1){20}}
\put(240,-1){\vector(0,1){20}}
\put(240,35){\vector(0,1){20}}
\put(235,60){$c_{p-2}$}

\put(280,60){$d_{p-2}$} 
\put(315,60){$\dots$}
\put(185,-10){$b_{0}$}
\put(235,-10){$c_{0}$}
\put(280,-10){$d_{0}$}
\put(315,-10){$\dots$}
\put(360,-10){$\dots$}
\put(185,26){$\dots$}
\put(235,26){$\dots$}
\put(280,26){$\dots$}
\put(315,26){$\dots$}
\put(360,26){$\dots$}
\put(400,-10){$\dots$}
\end{picture}
\end{equation}\\

 The vertical arrows denote action by the exterior derivative $d$, the horisontal arrows denote action by the de Rham differential $\delta$. The columns indicate the overlaps of open sets all the way to the maximal overlap. From left to right: $U_\alpha, U_{\alpha \beta}, U_{\alpha\beta\gamma},$ etc.  Small letters denote gauge parameters, e.g.,  $\Delta B_{p+1} =d b_p$ etc. Note that $H=dB$ is globally defined.

\section{Hull's $(C,G)$ model}

Hull's model, originally defined in six-dimensional spacetime, can be thought of as a generalisation of the square of an $(H,B)$ type model for a three-form $H$, although not all such models can  be interpreted this way.  In addition, in the supersymmetric case, the three-form $H$ is self-dual, and the Hull field-strength $G$ is a $(3,3)$-form self-dual with respect to both sets of indices. In the following we do not insist on duality constraints although this would be natural when supersymmetry is demanded. 

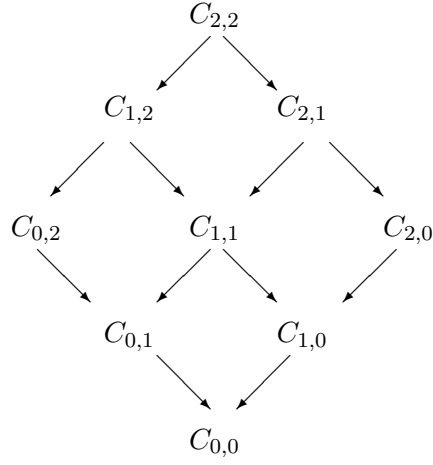
\be
\begin{picture}(400,200)
\put(202,180){$C_{2,2}$}
\put(210,175){\vector(-1,-1){20}}\put(215,175){\vector(1,-1){20}}
\put(170,145){$C_{1,2}$}\put(235,145){$C_{2,1}$}
\put(170,135){\vector(-1,-1){20}}\put(180,135){\vector(1,-1){20}}\put(245,135){\vector(-1,-1){20}}\put(255,135){\vector(1,-1){20}}
\put(135,100){$C_{0,2}$}\put(202,100){$C_{1,1}$}\put(275,100){$C_{2,0}$}\put(280,95){\vector(-1,-1){20}}
\put(145,95){\vector(1,-1){20}}\put(170,60){$C_{0,1}$}
\put(210,95){\vector(-1,-1){20}}
\put(215,95){\vector(1,-1){20}}\put(235,60){$C_{1,0}$}
\put(190,55){\vector(1,-1){20}}\put(240,55){\vector(-1,-1){20}}
\put(202,20){$C_{0,0}$}
\end{picture}
\la{4.13.b}
\ee

In this diagram $C_{2,2}$ is defined on each $U_{\a}$, $C_{1,2}$ and $C_{2,1}$ on each overlap $U_{\a\b}$ and so on down to $C_{0,0}$ which is defined on each quintuple overlap $U_{\a\b\c\d\e}$. The arrows indicate how the fields in different overlaps are related to fields at the next level. Thus one has

\bea
(\d C_{2,2})_{\a\b}:= C_{2,2 \a}-C_{2,2 \b}
=d_1 C_{1,2\a\b} + d_2 C_{2,1\a\b}
\eea
This then implies that
\be
d_1 (\d C_{1,2})_{\a\b\c}+ d_2(\d C_{2,1})_{\a\b\c}=0
\ee
where
\be
{(\d C_{1,2})}_{\a\b\c}:=C_{1,2\, \a\b} + C_{{1,2}\, \c\a} + C_{{1,2}\,\b\c}\ .
\ee
 This follows from  the first two lines of (4.6) below, because $d_1$ and $d_2$ are exterior derivatives acting on the first and second sets of indices satisfying

\be
(d_1)^2=(d_2)^2=d_1 d_2 + d_2 d_1=0\ .
\ee
From the diagram it is straightforward to read off the variations of the various $C$s. They are:

\bea
(\d C_{1,2})_{\a\b\c}&=&d_1 C_{0,2 \a\b\c} +d_2 C_{1,1\a\b\c}\nn\w1
(\d C_{2,1})_{\a\b\c}&=&d_1 C_{1,1\a\b\c} + d_2 C_{1,0\a\b\c}\nn\w1
(\d C_{0,2})_{\a\b\c\d}&=&d_2 C_{0,1\a\b\c\d} \nn\w1
(\d C_{1,1})_{\a\b\c\d}&=&d_1 C_{0,1 \a\b\c\d}+d_2 C_{1,0\a\b\c\d}\nn\w1
(\d C_{2,0})_{\a\b\c\d}&=&d_1 C_{1,0 \a\b\c\d}\nn\w1
(\d C_{0,1})_{\a\b\c\d\e}&=&d_2 C_{0,0 \a\b\c\d\e}\nn\w1
(\d C_{1,0})_{\a\b\c\d\e}&=&d_1 C_{0,0 \a\b\c\d\e}\nn\w1
d(\d C_{0,0})_{\a\b\c\d\e\f}&=&0 .
\eea
where at the last step $d=d_1=d_2$. Thus $(\d  C_{0,0})_{\a\b\c\d\e\f}$ is a set of constants defined on sextuple overlaps.

Note that the above diagram (3.1) is symmetrical about the vertical axis so that it essentially collapses to a vertical descent series, except for the middle
row where $C_{2,0}$ is the same as $C_{0,2}$, so that there are two fields here, $C_{1,1}$ and $C_{0,2}=C_{2,0}$. For this reason we shall consider the more general case of a non-symmetric $C_{p,q}$ field defined on $U_{\a}$, with field strength $G_{p+1,q+1}=d_1 d_2 C_{p,q}$. $C_{p,q}$ therefore corresponds to the two column Young tableau with $p$ and $q$ boxes in the first and second rows respectively.

\section{General $C_{p,q}$ model}

To illustrate the general case it will suffice to consider the simple example of $C_{3,2}$:

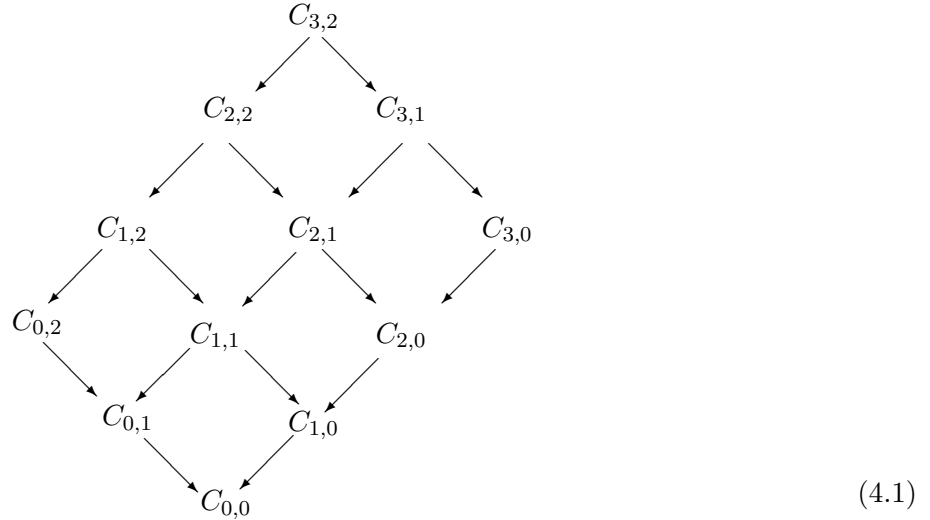
\be
\begin{picture}(400,200)
\put(202,180){$C_{3,2}$}
\put(210,175){\vector(-1,-1){20}}\put(215,175){\vector(1,-1){20}}
\put(170,145){$C_{2,2}$}\put(235,145){$C_{3,1}$}
\put(170,135){\vector(-1,-1){20}}\put(180,135){\vector(1,-1){20}}\put(245,135){\vector(-1,-1){20}}\put(255,135){\vector(1,-1){20}}
\put(130,100){$C_{1,2}$}\put(202,100){$C_{2,1}$}\put(275,100){$C_{3,0}$}\put(280,95){\vector(-1,-1){20}}
\put(132,95){\vector(-1,-1){20}}
\put(98,65){$C_{0,2}$}
\put(110,60){\vector(1,-1){20}}
\put(150,95){\vector(1,-1){20}}\put(165,60){$C_{1,1}$}
\put(132,29){$C_{0,1}$}
\put(148,24){\vector(1,-1){20}}
\put(169,-3){$C_{0,0}$}

\put(205,94){\vector(-1,-1){20}}
\put(215,95){\vector(1,-1){20}}\put(235,60){$C_{2,0}$}
\put(186,57){\vector(1,-1){20}}\put(236,54){\vector(-1,-1){20}}
\put(202,27){$C_{1,0}$}

\put(204,25){\vector(-1,-1){20}}

\put(165,58){\vector(-1,-1){20}}
\end{picture}
\la{4.13.b}
\ee

We can interpret (4.1) in several different ways

To begin with we can take $C_{3,2}$ to be defined on $U_{\a}$, the second level, $C_{2,2}$ and $C_{3,1}$ to be defined on $U_{\a\b}$, and so on, so that the the different levels are connected by the $\d$ operation as in the diagram (4.1) and equation $(4.2)$. This is similar to diagram (4.1) above, again with differentials $d_1$ and $d_2$ acting to the left and right respectively.

Alternatively, we can take the whole diagram to be defined on $U_{\a}$, so that all the nodes below denote sequences of gauge transformations  $\Delta$ ending at the top node. For example, $\D C_{3,2}=d_1 C_{2,2}+d_2 C_{3,1}$, $\D C_{2,2}=d_1 C_{1,2}+d_2 C_{2,1}, \D C_{1,2}=d_1 C_{0,2}+d_2 C_{1,1}$, and so on. There are also intermediate possibilities: for example, we can take $C_{3,2}$ to be defined on $U_{\a}$, $C_{2,2}$ and $C_{3,1}$ and all lower nodes on $U_{\a\b}$. In this case there are two overlapping sequences of descendants. 

\section{Form sector of $D=11$ supergravity}.

As is well-known, $D=11$ supergravity has a three-form potential $A_3$ with a four-form field strength tensor $F_4=d A_3$,  as well as the graviton and gravitino. One can also include its dual, $F_7=dA_6 + A_3 F_4$. This is a superfield so that duality applies to the leading bosonic compnent. The Bianchi identities are

\be
dF_4=0;~~~ d F_7= F_4^2,
\ee

\noindent In terms of potentials,

\be 
F_4=d A_3; ~~~F_7=d A_6 + A_3 F_4.
\ee

In the bosonic sector for these fields one can also consider the case where they are non-trivial with respect to an open covering $(U_\a)$ of the type considered above. For the three-form $F_3$ we then have a descent of the form $\d A_3=d B_2; \d B_2= dC_1; \d C_1= dD_0$ where the potentials are sequentially defined on single, double, triple and quadruple overlaps. Similarly for the seven-form $F_7=d A_6 + A_3 F_4$, we have potentials $A_6, B_5, C_4,\dots G_0$. In each of the overlaps there will be local gauge invariances; for example, the pair $(A_3, A_6)\rightarrow (B_2,B_5)\ldots (D_0,D_3)$ and thereafter to $E_2,F_1$ and $G_0$ for the $A_6$ sector only. So for $A_3$ the descent sequence ends at $D_0$ while for $A_6$ it continues to $G_0$ which is defined on seven-fold overlaps.

On each of the overlaps there will be local gauge transformations, so that, for example, 

\begin{align}
\Delta (A_3,A_6)&=(d a_2,d a_5-a_2 F_4)\ \rm {on}\ U_{\a}\nn\w1
\Delta (B_2,B_5)&=(d b_1,d b_4+b_1 F_4)\ \rm{on}\ U_{\a\b}
\end{align}
and so on. We will  have local gauge transformations on each overlap. For example, on $U_{\a}$ will have gauge parameters in a sequence $(a_2,a_5)\rightarrow (a_1,a_4)\rightarrow (a_0,a_3)$ and then down to 
$(-,a_2), (-,a_1)$ and $(-,a_0)$, where the double usages of the same symbol should not be confusing?, while on double overlaps we will have $(B_5,B_2)\rightarrow (b_4,b_1)$ etc down to $(b_0,-)$. The series continues with $(C_4,C_1)\rightarrow (c_0,0)$ on triple overlaps etc. 

We illustrate the situation ``doubling'' the diagram  \re{Hdiagram} with  $(p_1,p_2)=(3,6)$ .\\

\begin{equation}  
\hspace{-8cm}
\begin{picture}(500,205)

\put(115,212){$(F_4 ,F_{7})$}
\put(155,215){\vector(1,0){20}}

\put(170,180){$(A_3,A_6)$}
\put(174,145){$(a_{2},a_5)$}
\put(174,100){$(a_1,a_{4})$}
\put(174,60){$(a_0,a_{3})$}
\put(174,26){$(-,a'_{2})$}
\put(174,-10){$(-,a'_{1})$}
\put(174,-50){$(-,a'_{0})$}

\put(210,102){\vector(1,0){20}}
\put(210,182){\vector(1,0){20}}
\put(210,182){\vector(1,0){20}}

\put(190,115){\vector(0,1){20}}
\put(190,73){\vector(0,1){20}}
\put(190,35){\vector(0,1){20}}
\put(190,-1){\vector(0,1){20}}
\put(190,-40){\vector(0,1){20}}


\put(250,155){\vector(0,1){20}}
\put(250,115){\vector(0,1){20}}
\put(250,73){\vector(0,1){20}}
\put(250,35){\vector(0,1){20}}
\put(250,-1){\vector(0,1){20}}
\put(250,-40){\vector(0,1){20}}

\put(231,145){$(B_{2},B_5)$}
\put(234,100){$(b_1,b_{4})$}
\put(234,60){$(b_0,b_{3})$}
\put(234,26){$(-,b_{2})$}
\put(234,-10){$(-,b'_{1})$}
\put(234,-50){$(-,b'_{0})$}

\put(210,148){\vector(1,0){20}}
\put(210,62){\vector(1,0){20}}
\put(210,30){\vector(1,0){20}}
\put(210,-8){\vector(1,0){20}}
\put(210,-48){\vector(1,0){20}}

\put(190,185){\vector(0,1){20}}
\put(190,155){\vector(0,1){20}}
put(190,73){\vector(0,1){20}}

\put(215,148){\vector(1,0){20}}

\put(233,100){$(C_1,C_4)$} 
\put(237,60){$(c_0,c_{3})$}
\put(237,28){$(-.c_{2})$}
\put(237,-8){$(-,c_{1})$}
\put(237,-48){$(-,c'_{0})$}

\put(290,60){$(-,D_{3})$}
\put(292,28){$(-,d_{2})$}
\put(292,-8){$(-,d_{1})$}
\put(293,-48){$(-,d_{0})$}

\put(348,28){$(-,E_2)$}
\put(350,-48){$(-,e_0)$}
\put(350,-8){$(-,e_1)$}

\put(402,-8){$(-,F_1)$}
\put(404,-48){$(-,f_0)$}

\put(460,-48){$(-,G_0)$}

\put(328,62){\vector(1,0){20}}
\put(328,30){\vector(1,0){20}}
\put(328,-8){\vector(1,0){20}}
\put(328,-48){\vector(1,0){20}}

\put(215,102){\vector(1,0){20}}
\put(215,62){\vector(1,0){20}}
\put(215,30){\vector(1,0){20}}
\put(215,-8){\vector(1,0){20}}
\put(215,-48){\vector(1,0){20}}

\put(252,73){\vector(0,1){20}}
\put(252,115){\vector(0,1){20}}
\put(252,35){\vector(0,1){20}}
\put(252,-1){\vector(0,1){20}}
\put(252,-40){\vector(0,1){20}}

\put(270,102){\vector(1,0){20}}
\put(270,62){\vector(1,0){20}}
\put(270,30){\vector(1,0){20}}
\put(270,-8){\vector(1,0){20}}
\put(270,-48){\vector(1,0){20}}

\put(419,-40){\vector(0,1){20}}
\put(309,73){\vector(0,1){20}}
\put(309,35){\vector(0,1){20}}
\put(309,-40){\vector(0,1){20}}
\put(309,-1){\vector(0,1){20}}

\put(366,-40){\vector(0,1){20}}
\put(365,-1){\vector(0,1){20}}

\put(382,30){\vector(1,0){20}}
\put(382,-8){\vector(1,0){20}}
\put(382,-48){\vector(1,0){20}}

\put(438,-8){\vector(1,0){20}}
\put(438,-48){\vector(1,0){20}}

\put(365,35){\vector(0,1){20}}
\put(419,-40){\vector(0,1){20}}
\put(419,-1){\vector(0,1){20}}
\put(475,-40){\vector(0,1){20}}

\end{picture}
\end{equation}\\

\vspace{1.5cm}
\noindent
The vertical arrows denote action by the exterior derivative $d$, while the horizontal arrows denote the differences between forms defined on various overlaps, usually denoted by $\d$. Thus $({\d A_3})_{\a\b}= A_{3 \a}- A_{3\b}$, while $({\d B_2})_{\a\b\c}=B_{\a\b}+B_{\c\a}+B_{\b\c}$, and so on.
The columns indicate the overlaps of open sets all the
way to the maximal overlap. From left to right: $U_\alpha, U_{\alpha \beta}, U_{\alpha\beta\gamma},$ etc., all the way to the sevenfold overlap. Small letters denote local gauge
parameters, e.g., $\Delta A_3 = da_2$ etc.

The vertical arrows denote action by the exterior derivative, but there are correction terms for each of the second entries in each pair. For example, $\d A_{3\a\b}= d B_{2\a\b}$ while $\d A_{6\a\b}=d B_{5\a\b}-B_{2_{\a\b}} F_4$,  so that the second top arrow in the second column represents   $\Delta B_5= db_4+b_1F_4 $. There are similar correction terms in each of the vertical columns for the local gauge invariances on each overlap.\\

Note that, in addition, we could include a potential $A_{10}$ with  field strength $F_{11}=dA_{10}+A_3F_4^2$ and Bianchi identity
$dF_{11}=F_4^3$. 
\bigskip

\noindent
Note added: After the completion of this paper we became aware of a recent paper \cite{Giotopoulos:2026yno} which also discusses the form sector of $D=11$ supergravity.

\end{document}